\documentclass[aps,prl,floats,showkeys,
  twocolumn,showpacs,superscriptaddress,reprint]{revtex4-1}

\usepackage{graphicx}

\usepackage{amsmath}

\usepackage[latin1]{inputenc}
\usepackage[british]{babel}
\usepackage{epsfig}
\usepackage{graphics,dcolumn,bm,epic, eepic,fleqn,float}
\usepackage{amssymb,amsmath,multirow,rotate,color}

\begin{document}

\title{Social cohesion, structural holes, and a tale of two measures}


\author{V. Latora}
\affiliation{School of Mathematical Sciences, Queen Mary University 
  of London, E1 4NS London (UK).}

\author{V. Nicosia}
\affiliation{School of Mathematical Sciences, Queen Mary University 
  of London, E1 4NS London (UK).}

\author{P. Panzarasa}
\affiliation{ School of Business and Management, Queen Mary University
  of London, E1 4NS London (UK)}

\begin{abstract}
In the social sciences, the debate over the structural foundations of
social capital has long vacillated between two positions on the
relative benefits associated with two types of social structures:
closed structures, rich in third-party relationships, and open
structures, rich in structural holes and brokerage opportunities. In
this paper, we engage with this debate by focusing on the measures
typically used for formalising the two conceptions of social capital:
clustering and effective size. We show that these two measures are
simply two sides of the same coin, as they can be expressed one in
terms of the other through a simple functional relation. Building on
this relation, we then attempt to reconcile closed and open structures
by proposing a new measure, Simmelian brokerage, that captures
opportunities of brokerage between otherwise disconnected cohesive
groups of contacts. Implications of our findings for research on
social capital and complex networks are discussed.
\end{abstract}



\maketitle

\section{Introduction}
\label{intro}

A fundamental idea in the social sciences is that social capital
originates from social relations. As a result, social structure has
long been seen as playing a crucial role in sustaining or hindering a
wide range of performance-related outcomes, both at the individual and
collective levels
\cite{granovetter_1973,granovetter_2005,lin_2001,lin_2001a,lingo_2010}. However,
while social scientists tend to agree on the salience of social
structure, there is still controversy over which type of social
structure matters as a source of social capital. Over the years,
scholars have typically advocated two opposite types of structure: the
``closed'' and ``open'' structures. On the one hand, proponents of the
benefits of closed structures draw on the idea that social cohesion
fosters trust \cite{burt_1995,coleman_1990,reagans_2003,uzzi_1997} and
a sense of belonging \cite{coleman_1988}, sustains cooperative
behaviour \cite{coleman_1988,ingram_2000} and the enforcement of
social norms \cite{coleman_1988,gargiulo_2009,granovetter_2005}, and
facilitates the creation of a common culture \cite{nahapiet_1998}. On
the other, advocates of the benefits of open structures emphasise the
value that actors can extract from being located near structural holes
separating non-redundant contacts, and thus from acting as brokers
between otherwise disconnected others
\cite{burt_1992,burt_2005,burt_2010,lingo_2010,stovel_2012}.

This paper aims to draw on the interplay between these two alternative
conceptions of social capital, and engage with the ongoing debate over
the relative benefits associated with closed and open structures. The
relation between these two types of social structure will be explored
through a comparative analysis of the measures with which these
structures have traditionally been operationalised and formalised: the
\textit{clustering coefficient} and the \textit{effective size} of an
actor's local neighbourhood. In particular, we will show that it is
possible to derive a simple mathematical relation between the
clustering coefficient and the effective network size of a node. The
existence of this relation between the two measures, which have
originally and independently been introduced with the purpose of
formalising two different concepts, supports the idea that social
cohesion and structural holes are no more than the two sides of the
same coin. Both measures can indeed be expressed in terms of number of
links and number of triangles incident upon a node.

The question as to whether social capital stems from open or closed
structures will probably always remain a matter of debate. There are
cases where one is more interested in the benefits coming from closed
structures \cite{gargiulo_2009}, and other cases where, conversely, it
is more convenient to exploit the existence of open structures
\cite{stovel_2012}. However, our work highlights that, to characterise
the local structural properties of a node, it is equally informative
to measure the clustering coefficient or the effective size of the
node's local neighbourhood. Drawing on this, we will then use the
relation between clustering and effective size to develop a novel
measure for a generative mechanism of social capital that lies at the
interface between closed and open structures: \textit{Simmelian
  brokerage}. Being sensitive to variations in the position of links
across local networks of the same density, this measure can capture
opportunities of brokerage between otherwise disconnected groups of
densely interconnected nodes
\cite{dekker_2006,krackhardt_1998,krackhardt_1999,krackhardt_2002,tortoriello_2010,vedres_2010}. For
this reason, Simmelian brokerage can be seen as suitable for
formalising structures that lie at the interface between the closed
and open ones \cite{burt_2005,tortoriello_2010,vedres_2010}: it
captures the extent to which a node's local network is characterised,
on the one hand, by a combination of structural cleavages between
distinct groups of contacts, and on the other by a closed cohesive
structure within the boundaries of each group of contacts.

The paper is organised as follows. In Section \ref{soccap} we
introduce the concept of social capital, and offer a general overview
of the two main theoretical conceptions of its structural
foundations. In Section~\ref{sec:3} we review the definitions of
clustering coefficient and network effective size. In
Section~\ref{sec:4} we show that the two measures are linked through a
simple mathematical relation, based on which, in Section~\ref{sec:5},
we introduce our new measure of Simmelian brokerage. Finally, in
Section~\ref{sec:6} we extend the relation between clustering and
effective size to the case of weighted graphs, and sketch out a
definition of weighted Simmelian brokerage. The last Section will
summarise and discuss our main findings.

\section{Structural foundations of social capital}
\label{soccap}

The premise that seems to underpin most perspectives on social capital
is the idea that investments in social relations yield expected
returns in the marketplace, including the community, the economic,
financial, political, and labour markets~\cite{lin_2001,lin_2001a}. As
argued by Coleman, social capital can be characterised by two distinct
properties: it "inheres in the structure of relations between actors
and among actors", and ``like other forms of capital, [it] is
productive, making possible the achievement of certain ends that in
its absence would not be possible"~\cite[p.S98]{coleman_1988}.

Social scientists have long agreed on the salience of social structure
as a source of social capital
\cite{granovetter_1973,granovetter_2005,lin_2001,lin_2001a,lingo_2010}.
For instance, social structure has been seen as playing a pivotal role
in sustaining individuals' and organisations' efforts to create
value~\cite{gargiulo_2009,ingram_2000,mizruchi_2001,reagans_2001},
generate and transfer new
knowledge~\cite{centola_2007,hansen_1999,reagans_2003,tortoriello_2010,tortoriello_2012},
and enhance
creativity~\cite{ahuja_2000,brass_1995,fleming_2007,obstfeld_2005,perry-smith_2006,sosa_2011,uzzi_2005}.
As suggested by a number of scholars~\cite{granovetter_2005,lin_2001},
three main explanations can be offered as to why social structure
affects the outcomes of purposive actions. First, social structure can
facilitate or hinder the flow of information, and in so doing it also
impacts on its
quality~\cite{hansen_1999,reagans_2003,tortoriello_2012,uzzi_2005}.
Second, social structure can be seen as a source of reward and
punishment due to the effects that social relations have on the
internalisation and enforcement of social norms, including those
against free-riding~\cite{gould_1991,ingram_2000}. Third, social
structure nurtures and promotes the attainment of actors' trust,
reputation, social credentials, status, identity and recognition
through processes of third-party referrals and reinforcement of
interactions~\cite{lin_2001,lin_2001a,uzzi_1997}.

Despite the convergence on the explanatory relevance of social
structure, however, there is still controversy and debate over the
type of social structure that matters as a source of social
capital~\cite{aral_2011,baum_2012,burt_2005,gargiulo_2000,lin_2001,lin_2001a,reagans_2001}.
Theoretical conceptions of the structural foundations of social
capital vacillate between two positions that vary in their
understandings of the benefits associated with two opposite types of
social structure: ``closed'' and ``open'' structures. Arguments in
favour of each of these structures have been inspired by distinct rich
traditions in sociological theory. Both arguments, however, are
conceptually rooted in Simmel's seminal theoretical contributions on
the expansion of a dyadic relationship into a three-party relationship
(``\textit{Verbindung zu dreien}"), and the sociological significance
of the third element~\cite{simmel_2011}. Simmel argued that the
introduction of a third party fundamentally changes the social
dynamics of a dyadic tie: ``The appearance of the third party
indicates transition, conciliation, and abandonment of absolute
contrast (although, on occasion, it introduces
contrast)."~\cite[p.145]{simmel_2011}. Simmel's emphasis here is on
the two alternative functional roles the third party can play in the
triad: the ``non-partisan'' or mediator with the \textit{tertius
  iungens} (or ``the third who joins") orientation on the one
hand~\cite{obstfeld_2005}, and the broker with the \textit{tertius
  gaudens} (or ``the third who enjoys") orientation on the
other~\cite{burt_1992}.

\subsection{\em Closed structures and social cohesion}

Proponents of the benefits of closed structures typically build on
Simmel's~\cite{simmel_2011} \textit{tertius iungens} logic and
Coleman's~\cite{coleman_1988,coleman_1990} conception of social
capital predicated on the mechanism of social
cohesion~\cite{friedkin_2004}. Over the years, the Simmelian triad has
provided the theoretical backdrop against which scholars have
investigated the relational hypothesis that actors separated by one
intermediary are more likely to become connected with each other than
actors that do not share any common
acquaintance~\cite{davis_1970,davis_1971,holland_1970,holland_1971,luce_1949,wattsbook,ws98}.
At the macro level of a social system, the tendency of actors to forge
links locally within groups is conducive toward the creation of
cohesive social structures, organised into well-defined tightly knit
communities that are densely connected within but not across
boundaries~\cite{fortunato_2010,lambiotte_2009}. 

One of the most influential theories of social capital, advocated by
Coleman~\cite{coleman_1988}, is predicated precisely on the benefits
that actors accrue from being socially embedded within cohesive social
structures, rich in third-party relationships. Among the closure-based
sources of social capital are normative control and deviance
avoidance~\cite{burt_2005,granovetter_2005,lin_2001,lin_2001a}. More
generally, network closure enables the emergence and enforcement of
social norms by encouraging the internalisation of standards of
acceptable behaviour and facilitating the detection and punishment of
defective behaviour~\cite{ingram_2000,uzzi_1997}. In addition, it has
been documented that being part of a close-knit group engenders a
sense of belonging~\cite{coleman_1988}, fosters
trust~\cite{burt_1995,coleman_1990,reagans_2003,uzzi_1997},
facilitates the exchange of fine-grained, complex, tacit, and
proprietary information~\cite{hansen_1999,uzzi_1997}, enables the
creation of a common culture and the emergence of a shared
identity~\cite{nahapiet_1998}, and helps sustain a high level of
cooperation~\cite{coleman_1988,ingram_2000}.

Despite the benefits associated with social cohesion, the tendency of
individuals to cluster into densely connected communities also bears a
two-fold cost: local redundancy and social pressure. On the one hand,
the more an actor's contacts are connected with each other, the less
likely they are to take the actor closer to diverse sources of
knowledge and resources that the actor is not already able to
access~\cite{granovetter_1973}. Paucity of connections with new and
non-redundant social circles may create isolation and eventually
degrade social capital. This is the central argument of
Burt's~\cite{burt_1992} seminal contribution on the benefits
associated with occupying brokerage positions between otherwise
disconnected individuals or groups in a network. On the other hand,
above and beyond the redundancy of knowledge and resources, a cohesive
structure can still exert a negative influence on the connected actors
as a result of the social pressure favouring convergent thinking and
group consensus. As dense third-party relationships engender
reciprocal behaviour and sustain high degrees of similarity among the
actors, they are conducive toward the maintenance of the
\textit{status quo} rather than the exploration of novel paths leading
to divergent solutions~\cite{fleming_2007,sosa_2011}.

\subsection{\em Open structures and brokerage}

Both types of costs - redundancy and social pressure - associated with
social cohesion have inspired an alternative conception of social
capital, typically distilled into the proposition that there are
benefits actors can extract from participating in open structures that
are rich in cleavages and opportunities of
brokerage~\cite{burt_1992,burt_2005,burt_2010,lingo_2010,stovel_2012}.
At the heart of this conception of social capital lies
Simmel's~\cite{simmel_2011} characterisation of the role of
\textit{tertius gaudens} in a triad. While the non-partisan
\textit{tertius iungens} aims ``to save the group unity from the
danger of splitting up"~\cite[p.154]{simmel_2011}, the \textit{tertius
  gaudens} wishes to create or intensify discontinuities in the social
structure by forging or preserving unique ties to disconnected others.

The idea that social capital can originate from brokerage
opportunities associated with structural gaps has been explored most
thoroughly by Burt, who has perhaps contributed more than any other
sociologist in recent decades to examine the structural features and
performance implications of brokerage, especially in organisational
domains~\cite{burt_1992,burt_2004,burt_2005,burt_2010}. Burt defines a
structural hole as the ``separation between non-redundant contacts'',
``a relationship of non-redundancy between two contacts'', ``a
buffer'' that enables the two contacts to ``provide network benefits
that are in some degree additive rather than
overlapping''~\cite[p.18]{burt_1992}. Burt further identifies two
sources of the social capital that an actor can mobilise by acting as
the broker between contacts at the opposite sides of the hole:
information benefits and control benefits. On the one hand,
information benefits originate from the fact that, in open structures
rich in structural holes, connections tend to be
weak~\cite{granovetter_1973} and are likely to link people with
different ideas, interests and perspectives~\cite{burt_2004}. By
gaining exposure to a greater variance and novelty of information,
actors embedded in brokered structures will be creative and successful
in their endeavours~\cite{burt_2004,fleming_2007,sosa_2011}. On the
other, control benefits are related to the third party's ability to
gain an advantage by negotiating his or her relationships with
disconnected others and turning their ``forces combined against him
into action against one
another.''~\cite[p. 162]{simmel_2011}. Preserving and fostering
disunion between parties thus enable the actor standing near a
structural hole to extract social capital buried in the hole, by
playing the disconnected parties' demands and preferences against one
another.

\subsection{\em The trade-off between closed and open structures}

A number of empirical studies have attempted to reconcile the two
positions on social capital, and provide an integrative account of
social cohesion and
brokerage~\cite{aral_2011,fleming_2007,perry-smith_2006,rodan_2004,tortoriello_2010,vedres_2010}. Even
though the routes pursued to develop a unified conception of social
capital vary both theoretically and methodologically, scholars seem to
converge on the idea that the benefits originating from social
structure are contingent on a number of social, structural, and
environmental
conditions~\cite{aral_2011,fleming_2007,perry-smith_2006,rodan_2004},
and that a suitable combination of the two types of structure can
outperform each individual type in
isolation~\cite{reagans_2001,tortoriello_2010,vedres_2010}.

A substantial body of the literature has examined the trade-off
between social cohesion and brokerage by focusing on the interplay
between social structure and the attributes of the interacting
individuals, and suggesting that the benefits of either type of
structure - closed or open - are contingent upon such
attributes~\cite{perry-smith_2006,reagans_2001,rodan_2004}. In this
vein, for example, Fleming et al.~\cite{fleming_2007} have empirically
examined the mitigating effects exerted by individuals' attributes on
the benefits associated with brokerage. Their study suggests that,
while brokerage between otherwise disconnected collaborators makes all
individuals more likely to create new ideas, at the same time there
are marginal contingent positive effects of social cohesion on
generative creativity when individuals and their collaborators bring
broad experience, have worked for multiple organisations, and have
connections with external contacts. Similarly,
Perry-Smith~\cite{perry-smith_2006} has offered evidence suggesting
that connections to contacts with heterogeneous background mediate the
relationship between weak ties and creativity, and that there are
interaction effects between centrality and number of outside ties upon
creativity.

Another related line of investigation has suggested that an
appropriate combination of cohesion and brokerage opportunities can
provide individuals with the necessary redundant relationships as well
as access to non-redundant information that facilitate task execution
and enhance performance~\cite{burt_2005,podolny_1997,reagans_2001}. In
this vein, there have recently been attempts to address and resolve
the trade-off between closed and open structures by advocating a
conception of social capital that is contingent on the microstructural
context in which bridging ties are embedded. From this perspective,
Tortoriello and Krackhardt~\cite{tortoriello_2010} have argued that
brokerage is most beneficial when the the bridging tie is a Simmelian
one, namely a tie in which the parties involved are reciprocally and
strongly connected to each other as well as reciprocally and strongly
connected to at least one common third
party~\cite[p.~24]{krackhardt_1998}. Because the advantages
traditionally associated with open structures have been found to be
contingent upon the Simmelian nature of the bridging ties, this study
has provided empirical evidence in favour of an integrative account of
social capital, according to which individuals can extract benefits
not simply from structural holes or a cohesive neighbourhood in
isolation, but from a combination of both structural configurations.

More recent work has proposed a refined contingent conception of
social capital by recasting the trade-off between closed and open
structures in terms of the trade-off between ``channel bandwidth''
(i.e., tie strength) and "network diversity" (i.e., richness in
structural holes)~\cite{aral_2011}. The main argument is that, while
structural gaps remain sources of diverse information, the total
amount of useful novel information tends nonetheless to be positively
affected also by how strongly and frequently individuals interact with
one another. Strong relationships, characterised by frequent social
interactions, typically found in cohesive closed
structures~\cite{granovetter_1973,burt_1992}, are likely to sustain
the flow of a large volume of rich non-redundant information that, it
is claimed, ``tends to be more detailed, cover more topics, and
address more complex, interdependent
concepts''~\cite[p.~94]{aral_2011} than the information flowing in a
network rich in weak ties with less frequent interactions. However,
because the strength of ties tends to contract as the social structure
becomes richer in cleavages and brokerage opportunities, and thus more
diverse~\cite{burt_1992,granovetter_1973}, then a trade-off exists
between network diversity and tie strength as they produce
counterbalancing effects on the access to novel information. This
trade-off is resolved by regarding the relative benefits of network
diversity and tie strength as contingent on the social settings and
information environments in which individuals interact. In particular,
evidence has suggested that tie strength (i.e., closed structures)
trumps network diversity (i.e., open structures) as the topic space
becomes broader, information is frequently updated, and the overlap
between the information possessed by an individual's contacts becomes
larger~\cite{aral_2011}.

In this paper, we draw on these recent studies on social capital, and
contribute to the ongoing debate in a two-fold way. First, we
formalise the trade-off between closed and open structures by
proposing a functional relation between the measures with which these
two types of structure have traditionally been operationalised. Unlike
other studies~\cite[e.g.,][]{aral_2011,fleming_2007,tortoriello_2010},
we do not carry out an empirical investigation to test the relative
advantages of different structural configurations. By contrast, we
offer a rigorous and quantitative framework for a better understanding
of the trade-off between two concepts - cohesion and structural holes
- that have heretofore been compared to each other primarily, if not
exclusively, on intuitive grounds.

Our second contribution to the debate builds on the proposed
formalisation of the relationship between cohesion and structural
holes to offer a new measure - Simmelian brokerage - for detecting the
degree to which an individual's structural position lies at the
interface between a closed and an open structure. This measure is
inspired by, and is in qualitative agreement with, other studies that
have suggested the idea that social capital can originate
simultaneously from both social cohesion and structural
holes~\cite{reagans_2001,tortoriello_2010,vedres_2010}. However, it
differs from previous formalisations in two ways. First, unlike other
studies~\cite{vedres_2010}, we do not use clique percolation
methods~\cite{palla_2005} to uncover an overlapping community
structure and construct a group-level measure of group intersection
and multiple membership. Second, unlike other scholars, we do no rely
upon actor-level attributes (e.g., tenure)~\cite{reagans_2001} or
exogenously defined cross-boundary
relationships~\cite{tortoriello_2010} to detect network heterogeneity
and structural gaps in an individual's local neighbourhood. By
contrast, the novelty of our measure lies precisely in the fact that
it is defined at the node level and detects directly, based on the
node's local neighbourhood, the extent to which the node belongs to
multiple groups that are both tightly knit and disconnected from each
other. In this sense, Simmelian brokerage dovetails with the idea that
multiple group membership enables a node to extract social capital
from its underlying structure by blending social cohesion with
structural holes.

\section{Measuring social cohesion and structural holes}
\label{sec:3}

If social cohesion and structural holes have long represented two
distinct conceptual pillars, each underlying one of the two opposing
conceptions of social capital, they have also been formalised through
two distinct, and independently developed, measures: respectively, the
clustering coefficient and the effective size of a node's local
network. While clustering has typically been used for measuring the
extent to which a node is socially embedded within a closed cohesive
structure~\cite{davis_1970,holland_1970,luce_1949,ws98}, effective
size is a measure for detecting the non-redundancy of a node's
contacts, and therefore the degree to which the node's local
neighbourhood is rich in structural holes~\cite{burt_1992}. The
remaining of this Section is organised into two
parts. Section~\ref{clustering} will review the definition of node
clustering coefficient, and discuss an alternative measure for
cohesion, node \textit{local efficiency}. Section~\ref{subsec:} will
be devoted to the formalisation of effective size.

\subsection{\em Clustering coefficient and local efficiency}
\label{clustering}

Let us consider an unweighted undirected graph $G(V,L)$ with $N=|V|$
nodes and $K=|L|$ links, and let us focus on one of its nodes, node
$i$, with $i \in \{1,2,\ldots,N\}$. In order to measure the local
cohesion of node $i$, we define the subgraph $G_i$ induced in $G$ by
the set ${\cal N}_i$ of the neighbours of $i$. The node clustering
coefficient $C_i$ of node $i$ can then be defined as
\cite{wattsbook,ws98}:
\begin{equation}
\label{eq:clusteringcoeff_node}
  C_i = \left\{ \begin{array}{cc}
{      \frac{\mbox{ $K[G_i]$} }{\mbox{ $k_i(k_i-1)/2$}}
        } & $for $  k_i \ge 2 \\
    0     & $for $   k_i=0,1
       \end{array}
\right.
\end{equation}
where $K[G_i]$ is the number of links in $G_i$. The node clustering
coefficient $C_i$ indicates the probability that two neighbours of
node $i$ are connected by a link, and is properly normalised by
definition such that $0 \leq C_i \leq 1$. In fact, the value of $C_i$
in Equation~(\ref{eq:clusteringcoeff_node}) is the ratio between the
actual number of links $K[G_i]$ in the subgraph induced by the first
neighbours of $i$ and their maximum possible number, that is
${{k_i}\choose{2}}= {k_i(k_i-1)}/2$. Notice that $K[G_i]$ is also
equal to the actual number of triangles containing node $i$, while
${k_i(k_i-1)}/2$ is the number of open triads centred on $i$, which
corresponds to the maximum possible number of triangles containing a
node $i$ with $k_i$ links. Therefore, the node clustering coefficient
$C_i$ in Equation~(\ref{eq:clusteringcoeff_node}) can be alternatively
seen as the proportion of triads centred in $i$ that close into
triangles.

\begin{figure*}[!t]
\centering
\includegraphics[width=6in,angle=0]{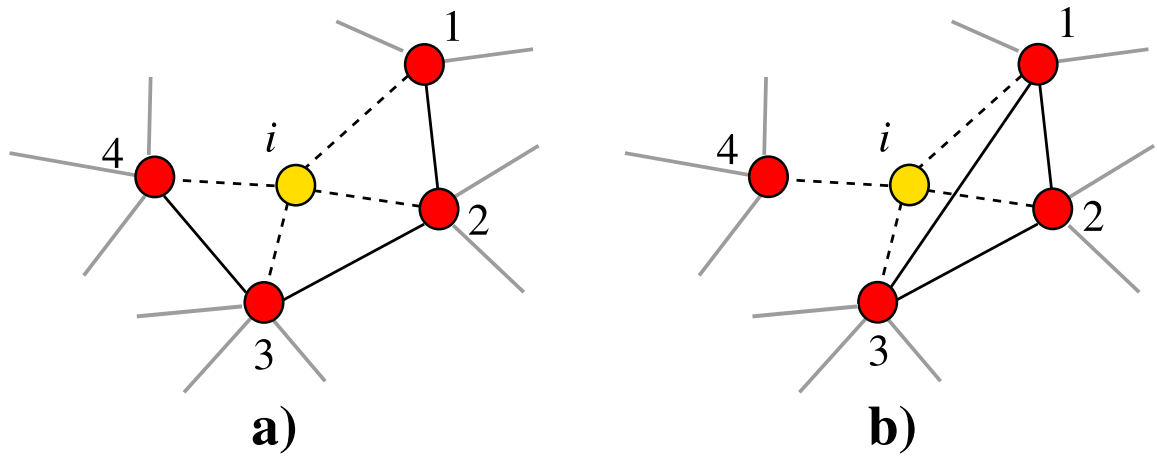}
\caption{The clustering coefficient of a node, as defined in
  Ref.~\cite{ws98}, measures the cohesiveness of the node's
  neighbourhood. Despite the different configurations of links, the
  clustering coefficient of node $i$, measured as the fraction of
  links among $i$'s neighbours over the total possible number of such
  links, is equal to $C_i=1/2$ in both graphs. Conversely, the local
  efficiency of node $i$ in graph a) is larger than in graph b).}
\label{fig:fig1}
\end{figure*}

As an example, let us consider the two graphs $G^a$ and $G^b$
shown in Figure~\ref{fig:fig1}. In both graphs, node $i$, coloured in
yellow, has degree $k_i=4$, so that the sets ${\cal N}^a_i$ and ${\cal
  N}^b_i$ of the neighbours of $i$ contain four nodes each. The four
neighbours of $i$, labelled as nodes $1$, $2$, $3$, $4$,
are shown as red circles, while the links connecting these
nodes to $i$ are shown as dashed lines.
The subgraph $G^a_i$ induced in $G^a$ by the set ${\cal N}^a_i$ of the
neighbours of $i$ has four nodes and three links shown as solid
lines. Similarly, the subgraph $G^b_i$ induced in $G^b$ by the set
${\cal N}^b_i$ contains four nodes and three links. Therefore, the
clustering coefficient of node $i$ is equal to $1/2$ in both cases. In
fact, the four nodes of the induced graphs can be connected to each
other through at most six links. In the figure, only three of these
potential links are present. Hence, $C_i = 3/6 = 1/2$. Notice that, in
general, the subgraph $G_i$ of the neighbours of node $i$ can be
unconnected. For instance, this happens for the case of graph b) in
Figure~\ref{fig:fig1}, where node $4$ is isolated in subgraph
$G^b_i$. However, this does not affect the mathematical definition of
the clustering coefficient $C_i$.  What is instead problematic is a
node $i$ which is itself an isolate in the graph, or has only one
link.  In this case, the ratio $2K[G_i]/ (k_i(k_i-1))$ is not
defined. The usual convention, in this case, is to set $C_i=0$ when
the degree of $i$ is either zero or one, as reported in
Equation~(\ref{eq:clusteringcoeff_node}).

\bigskip
An important feature of the clustering coefficient of node $i$, as
defined in Equation~(\ref{eq:clusteringcoeff_node}), is that it only
depends on the number of links in the subgraph $G_i$, and not on which
pairs of nodes are actually connected through such links in
$G_i$. Notice, for instance, that both $G_i^a$ and $G_i^b$ in
Figure~\ref{fig:fig1} have three links, but while $G_i^a$ is a line of
four nodes, $G_i^b$ consists of a triangle and an isolated node. For
this reason, here we discuss an alternative measure for the cohesion
of the neighbourhood of a node, the node local efficiency, that
enables the two cases shown in Figure~\ref{fig:fig1} to be clearly
differentiated.

In an unweighted undirected graph $G(V,L)$, the node local efficiency
of node $i$ is defined as the efficiency of the subgraph
$G_i$~\cite{lm01,lm03}, where the efficiency of a graph is the average
of the inverse of the distances between the nodes of the
graph. Therefore, the node local efficiency, $E_i$, of node $i$ can be
written as:
\begin{eqnarray}
\label{eq:localefficiency}
 E_i & = & E[G_i] = \frac{1} {k_i (k_i-1)} \sum_{\ell \in {\cal N}_i}
 \sum_{ \stackrel {m \in {\cal N}_i} {m\neq \ell} } \epsilon_{\ell m} =
 \nonumber\\ 
 & = & \frac{1} {k_i (k_i-1)} \sum_{\ell \in {\cal N}_i}
 \sum_{ \stackrel {m \in {\cal N}_i} {m\neq \ell} } \frac{1}{d_{\ell
     m}}
\end{eqnarray}
where $E[G_i]$ stands for the efficiency of graph $G_i$, while
$\epsilon_{\ell m}$ measures the reachability between node $\ell$ and
node $m$, and is set equal to the inverse of the distance $d_{\ell m}$
between the two nodes. Notice that distances between nodes are
evaluated on the graph $G_i$, and not on graph $G$. Moreover, the
local efficiency is properly normalised by definition, such that $0
\le E_i \le 1$. Therefore, it takes values in the same range as the
clustering coefficient.

By making use of Equation~(\ref{eq:localefficiency}), it is possible
to distinguish between the roles that node $i$ plays in the two graphs
$G^a$ and $G^b$ in Figure~\ref{fig:fig1}. If we calculate the
distances between the four neighbours of $i$ in $G^a_i$, we obtain:
$d_{12}=d_{23}=d_{34}=1$, $d_{13}= d_{24}= 2$ and $d_{14}=3$. Hence,
we have $\epsilon_{12}=\epsilon_{23}=\epsilon_{34}=1$,
$\epsilon_{13}=\epsilon_{24}=1/2$, and $\epsilon_{14}=1/3$, so that
the local efficiency of node $i$ in graph a) of Figure~\ref{fig:fig1}
is $E^a_i=13/18$. Conversely, if we consider $G^b_i$, we obtain:
$d_{12}=d_{23}=d_{13}=1$, $d_{14}= d_{24}=d_{34}=\infty$. Thus, we
have: $\epsilon_{12}=\epsilon_{23}=\epsilon_{13}=1$, $\epsilon_{14} =
\epsilon_{24} = \epsilon_{34} = 0$, and in this case the local
efficiency of node $i$ is $E^b_i= 1/2$, which is smaller than
$E^a_i$. Thus, even if node $i$ has the same clustering coefficient in
the two graphs, its local efficiency is different in the two cases.

The mathematical definition of efficiency we have adopted implies that
the efficiency of a graph with a fixed number of nodes becomes larger
as the number of links increases. And for graphs with the same number
of nodes and the same number of links, the efficiency depends on where
the links are actually located in each graph. In particular, the
efficiency of a chain of three links connecting four nodes is higher
than the efficiency of a triangle combined with an isolated node. The
reason for this is that, in the latter case, the presence of an
isolated node affects the overall reachability among nodes in the
system: notwithstanding the presence of a triad of connected nodes,
the isolated node is not reachable from the other ones.

\subsection{\em Effective size} 
\label{subsec:}

The original formalisation of the idea of structural holes was
proposed by Burt for weighted graphs \cite{burt_1992}. Therefore, here
we will first begin our analysis of measures for open structures with
the most general case of a directed weighted graph $G(V,L,W)$. We will
then restrict our focus to the particular case of undirected
unweighted graphs, and return to the general case of weighted graphs
in Section~\ref{sec:6}.

Let us indicate as $w_{ij} \ge 0$ the $(i,j)$ entry of the weighted
asymmetric matrix that describes the directed weighted graph
$G(V,L,W)$. As argued by Granovetter \cite{granovetter_1973}, the
weight of a link between any nodes $i$ and $j$ has the following
meaning: a high (low) value of $w_{ij}$ indicates a large (small)
amount of time, emotional intensity, intimacy, and reciprocal services
that characterise the link connecting node $i$ to node $j$. Among the
various measures introduced by Burt to detect and quantify the
presence of structural holes, a key role is played by the effective
size of a node's local network.

The effective size of node $i$'s network indicates the extent to which
each of the first neighbours of $i$ is redundant with respect to the
other neighbours, and can be expressed in terms of the two following
matrices: the \textit{transition matrix} $P$ and the \textit{marginal
  strength matrix} $M$. The entry $p_{i\ell}$ of matrix $P$ measures
the proportion of $i$'s network time and energy invested in the
relationship with node $\ell$, and is defined as~\cite{burt_1992}:
\begin{equation}
  p_{i\ell}  = 
    \frac{w_{i\ell} + w_{\ell i}}{\sum_m ( w_{im} + w_{mi} ) }  
\end{equation}
where $w_{i\ell} + w_{\ell i}$ is the sum of the weights of the two
links connecting $i$ to $\ell$, while $\sum_m ( w_{im} + w_{mi} )$ is
the total strength of node $i$. This is the sum of the out-strength,
$s^\text{out}_i = \sum_m w_{im}$, and the in-strength, $s^\text{in}_i
= \sum_m w_{mi}$, of $i$, i.e., the sum of the weights of all the
incoming and outgoing links incident upon $i$.  Notice that by
definition $0 \le p_{i\ell} \le 1 ~\forall i,\ell$, with $p_{i\ell}=0$
if there is neither a link from $i$ to $\ell$, nor a link from $\ell$
to $i$. Also, the transition matrix $P$ is stochastic: $\sum_\ell
p_{i\ell} = 1$.
The entry $ m_{j\ell}$ of the second matrix, the marginal 
strength matrix $M$, is defined as:
\begin{equation}
  m_{j\ell} = \frac{w_{j\ell} + w_{\ell j}}{\max_m ( w_{jm} + w_{mj} )
  } 
\end{equation}
Again, $0 \le m_{j\ell} \le 1 ~\forall j,\ell$, with
$m_{j\ell}=0$ if there is neither a link from $j$ to $\ell$, nor from
$\ell$ to $j$.
Notice that the two matrices $P$ and $M$ defined above are
non-symmetric. 

According to the definition given by Burt, the effective size ${\cal
  S}_i$ of node $i$'s network reads~\cite{burt_1992,borgatti97}:
\begin{equation}
 {\cal S}_i = \sum_{j \in {\cal N}_i} \left[1 - \sum_{\ell} p_{i\ell}
   m_{j\ell}\right] 
\label{eq:effective_size}
\end{equation} 
where ${\cal N}_i$ is the set of neighbours of $i$.  Excluding the
case where $i$ is an isolate, for which $\mathcal{S}_i\equiv 0$ by
definition, in general $ 1 \le {\cal S}_i \le k_i ~ \forall i$, that
is the effective size of node $i$'s network ranges from its smallest
value equal to $1$, when node $i$ belongs to a clique, to a maximum
value equal to the node degree $k_i$, when there are no links
$(j,\ell)$ connecting any two neighbours $j$ and $\ell$ within $i$'s
network, i.e. when $i$ is the centre of a star graph. In general, the
more redundant the neighbours of $i$ are, the smaller the value of
$\mathcal{S}_i$ is, and vice versa.

The expressions above largely simplify in the case in which the graph
is undirected and/or unweighted.  In fact, when the graph is
undirected, we have: $w_{i\ell} = w_{\ell i} ~ \forall i, \ell$. In
this case, the entries of the transition matrix $P$ and of the
marginal strength matrix $M$ read, respectively:
 \begin{equation}
  p_{i\ell}  =  \frac{w_{i\ell} }{\sum_m  w_{im}  }  = \frac{w_{i\ell} }{s^{\text{out}}_i} 
\end{equation}
and 
\begin{equation}
  m_{j\ell}  = \frac{w_{j\ell}}{\max_m  w_{jm}  }  
\end{equation}
Furthermore, if the graph is also unweighted, we have: $w_{i\ell} =
a_{i\ell} ~\forall i, \ell$, where $a_{i\ell}$ is equal to $1$ if
there is a link between $i$ and $\ell$, and to zero otherwise. This
implies that $\max_m w_{jm} = \max_m a_{jm} = 1$ for any node $j$ that
is not an isolated node. Consequently, the entries of the transition
matrix $P$ and of the marginal strength matrix $M$ reduce to:
 \begin{equation}
  p_{i\ell}  =  \frac{a_{i\ell} }{\sum_m  a_{im}  }  = \frac{a_{i\ell} }{k_i} 
 \label{eq:tm_unw}
\end{equation}
and 
\begin{equation}
\centering
  m_{j\ell}  = a_{j\ell} 
 \label{eq:msm_unw} 
\end{equation}
In this case, $P$ is the transition probability of a random walk on
the graph \cite{cover}, while the marginal strength matrix $M$
coincides with the adjacency matrix $A$.

\begin{figure*}[!t]
  \centering
  \includegraphics[width=6in]{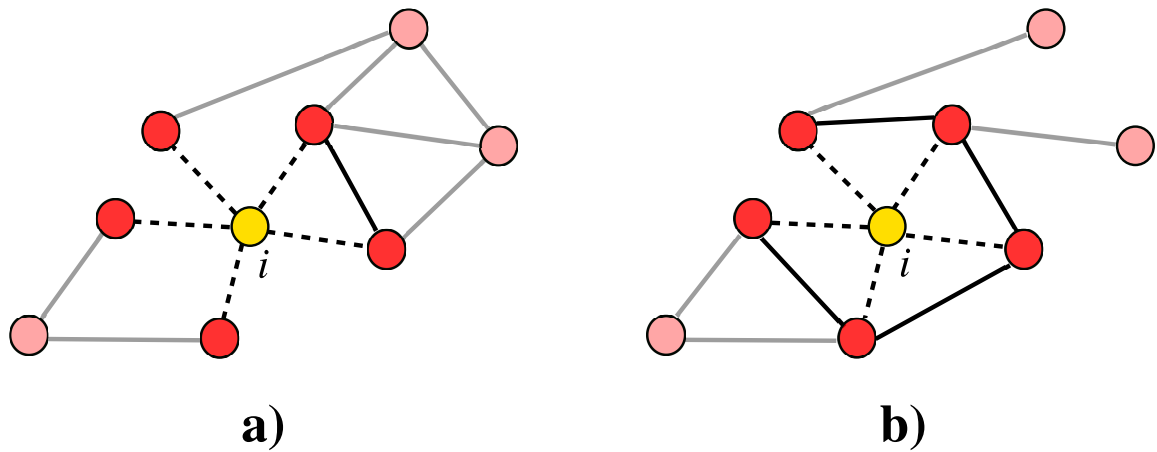}
  \caption{The effective size of a node's network, as defined in
    Ref.~\cite{burt_1992}, measures the lack of redundancy of the
    node's contacts. The neighbourhood of node $i$ in the first graph
    is less redundant than the one in the second graph. Consequently,
    the effective size, $\mathcal{S}^a_i = 23/5=4.6$, of node $i$ in
    graph $G^a$ is larger than the effective size,
    $\mathcal{S}^b_i=17/5=3.4$, of node $i$ in graph $G^b$.}
  \label{fig:effective_size}
\end{figure*}

Let us now consider the following example to illustrate the meaning of
Eq.~(\ref{eq:effective_size}). Figure~\ref{fig:effective_size} reports
two undirected unweighted graphs, $G^a$ and $G^b$, having the same
number of nodes, $N=8$, and the same number of links, $K=13$. In both
graphs, node $i$ is coloured in yellow and has $k_i=5$ links to its
neighbours, shown as dashed lines. The links connecting the neighbours
of $i$ are indicated by solid black lines. By visual inspection it can
be easily noticed that the neighbourhood of $i$ in graph $G^b$
contains more redundant links than in $G^a$. Indeed, if we evaluate
the effective size of node $i$ in graph $G^a$, we find that the
contribution of three out of the five neighbours in $i$'s network
towards the summation in Equation~(\ref{eq:effective_size}) is equal
to $1$. This is because each of these three nodes in ${\cal N}^a_i$
has no links to the other neighbours of $i$, and therefore is
non-redundant.
The contribution of each of the other two remaining neighbours in
$i$'s network towards the summation in
Equation~(\ref{eq:effective_size}) is equal to $1 - 1/k_i$. We finally
have: ${\cal S}^a_i= 1+1+1 +(1-1/5) + (1-1/5)= 3+ 8/5=23/5$, which is
a number larger than $3$, but smaller than the actual degree $k_i=5$
of the node. If we calculate the effective size of node $i$ in graph
$G^b$, we obtain: ${\cal S}^b_i= (1-1/5) + (1-2/5)+(1-2/5)+(1-2/5)+
(1-1/5)= 17/5 $. Because of the higher redundancy of the nodes in
${\cal N}^b_i$, this value is smaller than the effective size of node
$i$'s network in graph $G^a$.

\section{A simple relation between clustering and effective size}
\label{sec:4}

The two examples outlined in the previous Section already point to the
existence of a relation between the clustering coefficient and the
effective size of a node's network in an unweighted graph. In fact,
the values of $C_i$ and ${\cal S}_i$ both depend on the number of
triangles containing node $i$. The larger the number of triangles,
i.e. of closed structures involving node $i$, the larger the
clustering coefficient $C_i$. Conversely, the smaller the number of
triangles, the larger the number of open structures that can be
exploited by $i$, and thus the larger the effective size of node $i$'s
network. Therefore, we expect an inverse relation between $C_i$ and
${\cal S}_i$: the larger the clustering coefficient of a node, the
smaller the effective size of the node's network. In this Section, we
show that there is indeed an exact and simple relation between the two
measures. Based on this relation, it can then be argued that it is not
necessary to operationalise two distinct concepts (cohesion and
structural holes) and use two different measures (clustering and
effective size) to investigate two distinct sources of social capital,
when one source can be measured simply in terms of the other. More
specifically, either measure, together with the degree of the node, is
sufficient to quantify both the local cohesion and the structural
holes characterising the node's local network.

We first notice that the definition of node clustering coefficient
given in Equation~(\ref{eq:clusteringcoeff_node}) can be expressed in
terms of the adjacency matrix of the graph. In fact, the number of
links $K[G_i]$ in graph $G_i$ can be easily calculated from the
adjacency matrix by observing that $(A^3)_{ij} = \sum_{\ell ,m}a_{i
  \ell} a_{\ell m} a_{mj}$ is equal to the number of walks of length
$3$ connecting node $i$ to node $j$. In particular, by setting $i=j$,
the quantity $\sum_{\ell ,m} a_{i\ell} a_{\ell m} a_{mi}$ denotes the
number of closed walks of length $3$ from node $i$ to itself. This is
twice the number of triangles containing node $i$. The generic
triangle containing node $i$ and the two nodes $l$ and $m$ is made of
the two links connected to node $i$, namely $(i,\ell)$ and $(m,i)$,
and of the link $(\ell, m)$ that belongs to $G_i$. Since the link
$(\ell, m)$ appears twice, namely in the closed walk $(i,\ell,m,i)$
and in the closed walk $(i,m,\ell,i)$, the number of links $K[G_i]$ is
given by:
\begin{equation}
   K[G_i] = \frac{1}{2}\sum_{j,m}a_{ij}a_{jm}a_{mi}. 
\end{equation}
Notice that this is the numerator of
Equation~(\ref{eq:clusteringcoeff_node}), so that we can express the
local clustering coefficient of node $i$ as:
\begin{equation}
\label{eq:clusteringcoeff_node_A}
  C_i = \left\{ \begin{array}{cc}
    {  \frac{\mbox{ $ \sum_{j,\ell}a_{ij}a_{j\ell}a_{\ell i}$}}
      {\mbox{ $k_i(k_i-1)$   }}    
        } & $for $  k_i \ge 2 \\
    0     & $for $   k_i=0,1
       \end{array}
\right.
\end{equation}

Let us now consider the effective size of node $i$'s network. When the graph is
undirected and unweighted, the entries of the transition matrix $P$
and of the marginal strength matrix $M$ read, respectively, as in
Equations~(\ref{eq:tm_unw}) and (\ref{eq:msm_unw}). Consequently, the
effective size ${\cal S}_i$ of node $i$'s network can be written as:
\begin{eqnarray}
  \label{avvicinamento}
        {\cal S}_i & = & \sum_{j} a_{ij}\left[1 - \sum_{\ell} p_{i\ell}
          m_{j\ell}\right] = \nonumber\\
        & = & k_i - \sum_j a_{ij} \sum_{\ell} p_{i\ell}
        m_{j\ell} = \nonumber \\
        & = & k_i - \sum_j \sum_{\ell} a_{ij}
        \frac{a_{i\ell}}{k_i} a_{j\ell} = \nonumber\\
        & = & k_i - \frac{1}{k_i} \sum_{j}
        \sum_{\ell} a_{ij} a_{j\ell} a_{\ell i}
\end{eqnarray} 
If we plug the expression for the clustering coefficient of node $i$ in  
Equation~(\ref{eq:clusteringcoeff_node_A}) into Equation~(\ref{avvicinamento}), 
we obtain:
\begin{equation}
  \label{formulamagica}
        { \cal S}_i = k_i -(k_i-1) C_i
\end{equation} 
This is an exact relation that connects three measures at the node
level: effective size, clustering, and degree.  For instance, we can
use the relation to obtain the effective size of a node's network by
measuring the clustering coefficient of the node. Since for the two
graphs in Figure~\ref{fig:effective_size} we have $C_i^a= 1/10$ and
$C_i^b= 4/10$, by using Equation \ref{formulamagica} we obtain: ${
  \cal S}^a_i = 5 - 4/10= 23/5$ and ${ \cal S}^b_i = 5 - 16/10= 17/5$,
in perfect agreement with the values obtained by using Definition
\ref{eq:effective_size}. More generally,
Equation~(\ref{formulamagica}) provides a formalisation of the fact
that structural holes and social cohesion are indeed the two faces of
the same coin. The presence of structural holes, measured by the
effective size of a node's network, depends only on the degree of the
node and on the social cohesion of the node's local network, as
measured by the node clustering coefficient. Conversely, the
clustering coefficient of a node is uniquely determined by the degree
of the node and by the effective size of its network.

By definition, the effective size of a node's network can take values
between zero (if the node is an isolate) and the degree of the node,
while the clustering coefficient varies from zero to one. To make the
two quantities comparable, we can normalise the definition of
effective size. To this end, we can define the normalised effective
size of node $i$'s network, ${ \cal S}^{\prime}_i$, dividing the
effective size ${ \cal S}_i$ by its maximum possible value $k_i$,
namely: ${ \cal S}^{\prime}_i = { \cal S}_i / k_i$. If the node is an
isolate, we set ${ \cal S}^{\prime}_i = 0$.  When both measures are
normalised in $[0,1]$, their relation reads:
\begin{equation}
  \label{formulamagica_norm}
 { \cal S}^{\prime}_i       = 1 - \frac{k_i-1}{k_i} C_i
\end{equation} 
which, for nodes with large degree, is well approximated by: 
\begin{equation}
  \label{formulamagica_norm_approx}
 { \cal S}^{\prime}_i \simeq 1 - C_i
\end{equation} 
This equation indicates that the clustering coefficient and the
normalised effective size are indeed two complementary measures that
can be defined one in terms of the other. As a result, this relation
cautions against using both measures simultaneously for detecting
sources of social capital. For instance, the inclusion of both
clustering and effective size as covariates in a multivariate
regression model would inevitably entail problems of multicollinearity
due to the linear relation found between the two measures.

Drawing on this relation, we can express in terms of effective size
many of the results obtained for the clustering coefficient in real
networks. In particular, it has been found that in many networks the
clustering coefficient of a node scales with the degree of the node as
$k^{-\omega}$, where $0\le\omega\le
1$~\cite{Vazquez2002,Ravasz2003}. This means that high-degree nodes
tend to have a relatively smaller clustering coefficient than
low-degree nodes. Consequently, in such networks effective size will
increase with $k$, so that higher-degree nodes will exhibit a higher
effective size than lower-degree nodes. The reason for this is
precisely the mirror image of the argument typically proposed to
explain the inverse relation between clustering and degree: any two
neighbours of a large-degree node are more likely not to be directly
connected with each other than any two neighbours of a low-degree
node. Moreover, it is possible to use any model of random networks
with a tunable degree-dependent clustering
coefficient~\cite{Holme2002,Ravasz2003,Vazquez2003,Serrano2005} to
construct a random network with a fixed distribution of effective
size. In turn, such random network can be used as a null model to
assess the statistical significance of the correlation, measured at
the node level, between effective size on the one hand, and other
structural properties or performance-related outcomes, on the other.

\section{Reconciling social cohesion and structural holes: Simmelian brokerage}
\label{sec:5}

As discussed in Section \ref{clustering}, local efficiency is a
generalisation of the node clustering coefficient, in that it measures
the extent to which the neighbours of node $i$ would reach each other
if $i$ were removed from the network. Unlike the clustering
coefficient, local efficiency allows us to distinguish between cases
where the subgraphs $G_i$ have the same number of nodes and the same
number of links, but are topologically different, as was the case of
the two examples in Figure~\ref{fig:fig1}. In this Section, we show
that local efficiency can be employed effectively to develop a new
measure for brokerage that lies at the interface between clustering
and effective size. Like clustering and effective size, this measure
will be sensitive to structural gaps in a node's local
network. However, unlike clustering and effective size, it will also
be sensitive to variations in the position of links across local
networks of the same density. For this reason, the measure is capable
of capturing brokerage opportunities among otherwise disconnected
socially cohesive groups of nodes.

We begin by observing that, if the local efficiency of node $i$ is
small, then $i$ plays an important role in enabling and facilitating
reachability among its neighbours. In this case, $i$ acts as a broker
among its neighbours, since the removal of $i$ would disconnect many
pairs of nodes in $i$'s neighbourhood, or would inevitably deteriorate
the ability of these nodes to reach each other. Conversely, if the
local efficiency of $i$ is large, then the nodes in $i$'s
neighbourhood would still be able to reach each other even without the
intermediary role of $i$, and as a result they would barely be
affected by the removal of $i$. In this case, $i$ plays a negligible
brokerage role. More generally, the higher the local efficiency of a
node, the fewer the opportunities a node has to act as a broker, and
vice versa. Based on this observation, on the relation in
Equation~(\ref{formulamagica}), and on the fact that the clustering
coefficient and the normalised effective size range in the same
interval $[0,1]$, here we introduce the following measure for local
brokerage $\mathcal{B}_i$ of node $i$:
\begin{equation}
\label{formulaproposta}
  { \cal B}_i = k_i -(k_i-1) E_i
\end{equation} 

In qualitative agreement with
Krackhardt's~\cite{krackhardt_1998,krackhardt_1999,krackhardt_2002}
idea of \textit{Simmelian ties} as ties embedded in cliques, we
propose to call this measure Simmelian brokerage. Our choice is
motivated by the fact that the measure is indeed sensitive to the
extent to which a node acts as a broker between Simmelian ties or,
alternatively, between otherwise disconnected groups of densely
connected nodes. This is the case of a node that is a member of
different cliques, and thus acts as the intermediary between two or
more disconnected sets of Simmelian ties, rich in third-party
relationships~\cite{burt_1998,dekker_2006,tortoriello_2010,vedres_2010}.
The definition of Simmelian brokerage is similar to that of effective
size, with the only difference that the clustering coefficient $C_i$
of node $i$ is replaced by the node local efficiency $E_i$. According
to Equation~(\ref{formulaproposta}), when the degree of a node is
fixed, an increase in the value of local efficiency corresponds to a
decrease in the value of Simmelian brokerage, and vice versa.

\begin{figure*}[!t]
  \centering \includegraphics[width=6in]{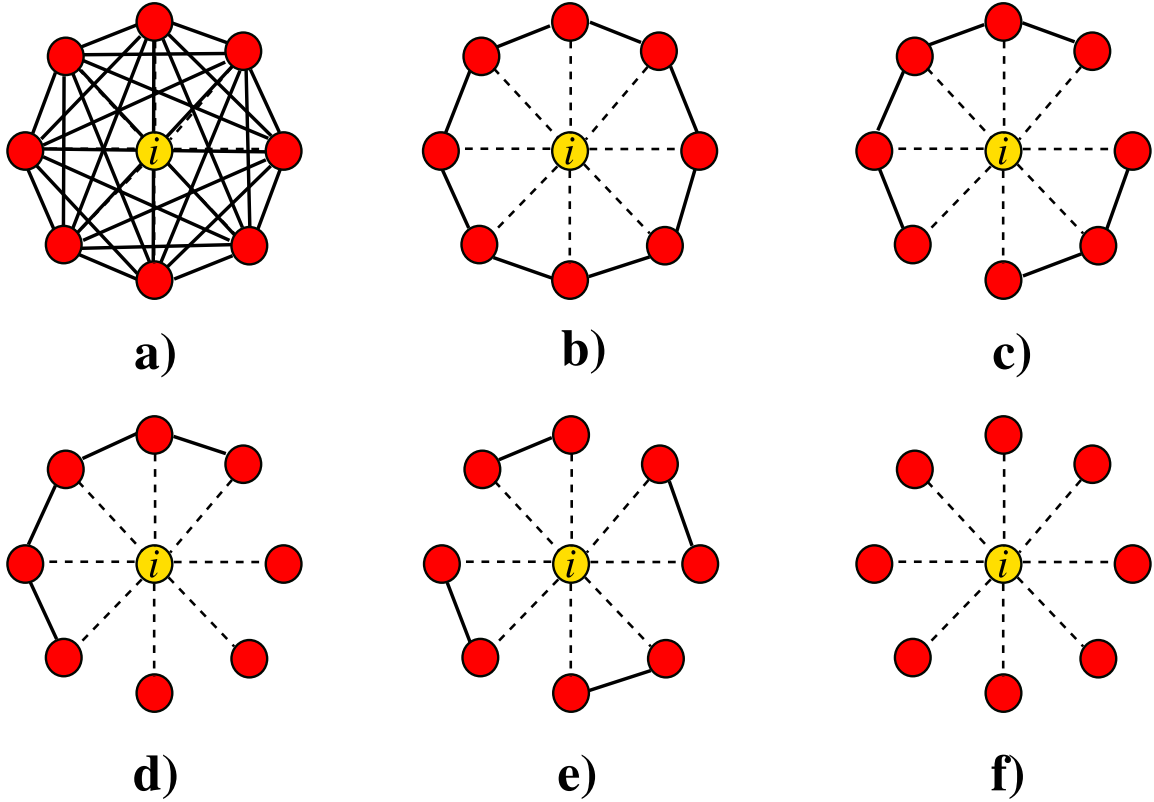}
  \caption{The extent to which a node acts as a Simmelian broker
    depends on the number and configuration of links in its
    neighbourhood. The six graphs correspond to six values of
    $\mathcal{B}_i$, in increasing order: a) $1.0$, b) $4.08$, c)
    $5.83$ d) $6.46$, e) $7.0$, f) $8.0$. When the neighbourhood of
    $i$ is a clique (graph a)), the Simmelian brokerage of $i$ is
    negligible, and $i$ is practically superfluous. As soon as the
    neighbourhood becomes sparser and more structural holes appear,
    node $i$ acquires higher values of Simmelian brokerage. In graphs
    d) and e), node $i$ has the same clustering coefficient (and thus
    the same effective size), but the value of Simmelian brokerage of
    $i$ is higher in e) than in d) because in e) $i$ intermediates
    among four cohesive groups of nodes, whereas in d) $i$
    intermediates between one group of four loosely connected nodes
    and three otherwise isolated nodes. Opportunities of brokerage are
    maximised in the limiting case of a star graph in which node $i$
    intermediates among eight otherwise disconnected contacts (graph
    f)).}
  \label{fig:fig_broker}
\end{figure*}

To shed light on the relation between effective size and Simmelian
brokerage, we now discuss a number of examples of brokerage
opportunities in unweighted undirected
graphs. Figure~\ref{fig:fig_broker} shows six graphs, each with $N=9$
nodes, but with a different number and configuration of links. A first
inspection of the figure makes it immediately clear that the central
node $i$, indicated in yellow, has a different brokerage role in each
of the six graphs. Graph a) is a clique, i.e. a complete graph in
which each node has a link to each of the other nodes. This structure
is characterised by high redundancy -- indeed the maximum possible
redundancy among all the graphs with the same number of nodes -- due
to the presence of the maximum possible number of links in the
graph. In this case, the local efficiency of node $i$ is equal to
$E^a_1=1.0$, and for Simmelian brokerage we obtain the smallest
possible value, $\mathcal{B}^a_i=8 - 7\times 1.0 = 1.0$. Such a small
value is consistent with the relatively negligible role that $i$ has
in facilitating reachability among its neighbours: if $i$ were removed
from the graph, not only would its neighbourhood remain connected, but
each of its contacts would still have a direct connection with each
other. For the overall network, node $i$ is thus a redundant contact.

Graph b) is a wheel graph, where the neighbours of node $i$ are
arranged in a cycle. As in the case of a clique, here the
neighbourhood of node $i$ remains connected even when $i$ is removed
from the graph, so that $i$ can be considered somehow
redundant. However, while some pairs of nodes in the induced graph
have distance equal to $1$, the majority of pairs of nodes are at
distance $2$, $3$ or $4$, so that the corresponding local efficiency
is $E^b_i\simeq 0.559$ and node $i$'s Simmelian brokerage is equal to
$\mathcal{B}^b_i\simeq 4.083$. The relatively higher value of $i$'s
Simmelian brokerage in graph b) than in graph a) reflects the more
central role played by $i$ in b) than in a) in facilitating
reachability among its neighbours. This is also reinforced by the fact
that, when $i$ is removed from the graph, the average distance among
its neighbours increases from $\sim 1.714$ (in graph a)) to $\sim
2.286$ (in graph b)).

From a visual inspection, we would also expect the Simmelian brokerage
of node $i$ to increase from graph b) to graph c) due to an increase
in structural gaps between distinct groups of nodes. In graph c), node
$i$ does not only intermediate between already connected nodes, but it
brokers between different groups of nodes that would otherwise remain
disconnected. In this case, the local efficiency of $i$ is
$E^c_i\simeq 0.309$ and, as expected, the corresponding Simmelian
brokerage of $i$ is higher than in graph b), and is equal to
$\mathcal{B}^c_i\simeq 5.83$.

Graphs d) and e) have the same number of links $K=4$, but with a
different configuration. Interestingly, the effective size of node
$i$'s network in the two graphs is the same, ($S_i=7.0$), since in
both graphs the clustering coefficient of $i$ is equal to
$C_i=1/7$. However, a removal of $i$ causes a more significant damage
in graph e) than in graph d), since the intermediary role of $i$ is
more crucial in e) than in d). Moreover, in graph e) node $i$ is
affiliated with four distinct cohesive groups of connected contacts,
whereas in graph d) node $i$ intermediates between three isolated
contacts and one loosely connected group. In this sense, in graph e)
node $i$ spans more structural holes between Simmelian ties than in
graph d). Our measure of Simmelian brokerage does indeed capture this
difference. In fact, if we removed $i$ from graph d), then $10$ out of
$28$ pairs among $i$'s neighbours (namely all pairs involving the five
nodes in the group) would still remain reachable. In this case, the
value of Simmelian brokerage of $i$ is $\mathcal{B}^d_i\simeq
6.46$. Conversely, the removal of $i$ in e) would produce a more
serious damage to the network, since only $4$ pairs of neighbours of
$i$ over $28$ would remain reachable. In this case, the value of
Simmelian brokerage is equal to $\mathcal{B}^e_i=7.0$.

Finally, graph f) is a star, so that by removing node $i$ no pair of
its neighbours would remain reachable any longer. In this limiting
case, the value of local efficiency of node $i$ is equal to zero, and
the one of Simmelian brokerage, as well as of effective size, is equal
to the node degree, namely $\mathcal{B}^f_i=8.0$.
In general, if two nodes have the same local efficiency, the one
having the higher degree has a higher Simmelian brokerage. This is due
to the fact that the removal of a high-degree node could potentially
leave a higher number of pairs of nodes disconnected, and could
therefore cause a more substantial damage to the network than the
removal of a node characterised by a relatively low degree.

\section{Weighted graphs: Effective strength and weighted Simmelian brokerage}
\label{sec:6}

In this Section, we briefly discuss how the measures of effective size
and Simmelian brokerage can be intuitively extended to the case of
weighted graphs. First, we notice that, in the more general case of
undirected weighted graphs, the importance of node $i$ can be measured
through its total strength $s_i=\sum_{j}w_{ij}$, in addition to the
degree $k_i$. Consequently, it is reasonable to extend the measure of
effective size to weighed graphs by defining the \textit{effective
  strength} of a node as follows:
\begin{eqnarray}
  \mathcal{S}^{w}_i & = & \sum_{j} w_{ij} \left[1 - \sum_{\ell} p_{i\ell}
    m_{j\ell}\right] = \nonumber\\
  & = & s_{i} - \sum_{j}\sum_{\ell}w_{ij}p_{i\ell}
  m_{j\ell}
  \label{eq:effective_strength}
\end{eqnarray}
Like effective size for unweighted graphs, effective strength measures
the extent to which the neighbourhood of a node in a weighted graph is
redundant. However, unlike effective size, it properly takes into
account the weights of links, and thus captures variations in the
investment (e.g., time, energy) that a node $i$ makes in each of its
neighbours. Since the quantity in square brackets in
Equation~(\ref{eq:effective_strength}) is multiplied by $w_{ij}$, then
a neighbour $j$ of node $i$'s for which $w_{ij}$ is relatively small
has little impact on the effective strength of $i$. Conversely, if
$w_{ij}$ is relatively large, then node $j$ can substantially
influence the effective strength of $i$.
It is easy to verify that, if the graph is unweighted and undirected,
the effective strength in Equation~(\ref{eq:effective_strength})
reduces to the effective size, as defined in
Equation~(\ref{eq:effective_size}).

In principle, starting from Equation~(\ref{eq:effective_strength}), it
should be possible to derive an exact relation between the effective
strength and the weighted clustering coefficient of a node, as we did
in Section~\ref{sec:4} for the unweighted case, and to extend to
weighted graphs the measure of Simmelian brokerage proposed in
Equation~(\ref{formulaproposta}). However, while the definition of
clustering coefficient in unweighted graphs reported in
Equation~(\ref{eq:clusteringcoeff_node}) is widely accepted and
undisputed, there exist more than one manner to define the clustering
coefficient of a node in a weighted graph~\cite{opsahl_2009}. Indeed,
different measures for the weighted clustering coefficient have been
proposed in the literature. Among those, the following four are the
most popular ones:

\medskip
\begin{alignat*}{2}
    C^{w}_i &= C^{B}_i = \frac{1}{s_i(k_i-1)}\sum_{j,\ell} \frac{
      w_{ij}+w_{i\ell}}{2} a_{ij}a_{j\ell}a_{\ell i} && \quad
    \\ C^{w}_i &= C^{O}_{i} =
    \frac{2\sum_{j,\ell}(w_{ij}w_{j\ell}w_{\ell i})^{1/3}}{k_i(k_i-1)}
    && \quad \\ C^{w}_i &= C^{Z}_{i} = \frac{\sum_{j\neq
        i}\sum_{j\neq\ell,\ell\neq i} (w_{ij}w_{j\ell}w_{\ell
        i})}{(\sum_{j\neq i} w_{ij})^2 - \sum_{j\neq i} w_{ij}^2} &&
    \quad \\ C^{w}_i &= C^{H}_i =
    \frac{\sum_{j,\ell}w_{ij}w_{j\ell}w_{\ell i}}{\max(w)
      \sum_{j,\ell}w_{ij}w_{\ell i}} && \quad 
\end{alignat*}

\medskip
\noindent
which have been respectively defined by Barrat et
al.~\cite{Barrat2004}, Onnela et al.~\cite{Onnela2005}, Zhang and
Horvath~\cite{Zhang2005} and Holme et al.~\cite{Holme2007}. Notice
that all these measures are essentially based on the same idea: the
clustering of node $i$ is measured by means of the sum of the weights
of the closed triads incident on $i$. Nevertheless, each measure
differs from the others in the choice of the weight assigned to each
triad and in the normalisation introduced to guarantee that $C^w_i$
takes values in $[0,1]$. A discussion of the different definitions of
the weighted clustering coefficient is beyond the scope of the present
paper (the authors of Ref.~\cite{Saramaki2007} have carried out a
thorough analysis of these measures and a comparison of their
properties). In a similar way as in the case of unweighted graphs, for
which the relation between effective size, clustering and degree of a
node is given by Equation~(\ref{formulamagica_norm}), for each of the
four definitions of weighted clustering it is possible to find a
corresponding functional relation to obtain the effective strength of
a node if one knows the value of its weighted clustering
coefficient. In general, if we assume that this functional relation is
mediated not only by the degree $k_i$ of node $i$ but also by the node
strength $s_i$, we can write:

\medskip
\begin{equation}
  \mathcal{S}^{w}_i = F(k_i, s_i, C^{w}_i)
  \label{eq:size_w}
\end{equation}

\medskip\noindent where the form of $F(k_i, s_i, C^{w}_i)$ depends
only on the chosen definition of clustering coefficient.  Following
the same logic described in Section~\ref{sec:5}, we notice that each
version of the weighted clustering coefficient induces a different
definition of weighted Simmelian brokerage $\mathcal{B}^{w}_i$. As we
did for the case of unweighted graphs, where the Simmelian brokerage of
a node was obtained by substituting $E_i$ for $C_i$ in
Equation~(\ref{formulamagica}), we define the weighted Simmelian
brokerage of node $i$ induced by a given definition of weighted
clustering as follows:

\begin{equation}
  \mathcal{B}^{w}_i = F(k_i, s_i, E^{w}_i)
  \label{eq:broker_w}
\end{equation}

\medskip\noindent where $E^{w}_i$ is the local efficiency of node $i$
in the weighted graph, which is measured considering the weighted
distances $d^{w}_{j\ell}$ instead of the topological distances
$d_{j\ell}$. In other words, for a given formulation of weighted
clustering, the Simmelian brokerage of a node is obtained by replacing
the weighted clustering coefficient $C^{w}_i$ with the local
efficiency $E^{w}_i$ in the function $F(k_i,s_i,C^{w}_i)$ that relates
effective strength to the weighted clustering coefficient.

\section{Conclusions}

Graphs are an invaluable mathematical tool for examining the topology
and evolution of social structures, and graph measures have
contributed to the operationalisation and formalisation of fundamental
sociological concepts as well as to the development of social
theories. Among these measures, clustering and effective size have
played a pivotal role in the debate that, over the last few decades,
has been concerned with the types of social structures that matter as
sources of social capital
\cite{aral_2011,baum_2012,burt_2005,fleming_2007,gargiulo_2000,lin_2001,lin_2001a,reagans_2001,tortoriello_2010}. In
this paper, our contribution to this debate began by reviewing the two
measures, clustering and effective size, typically associated with two
opposing types of social structure, the closed and open structure
respectively. We then clarified the relationship between these two
measures, and found that they are indeed connected through a simple
mathematical relation. While so far the two measures have been related
to each other primarily at a conceptual level and on intuitive grounds
\cite{burt_1992}, in this paper we provided a formal framework in
which one measure can be expressed in terms of the other.

The study of formal relations between different graph measures can
help unveil the intimate connections between already existing, and
apparently unrelated, sociological concepts and, at times, even lead
to the development of new concepts and measures. This indeed describes
the trajectory that brought us from a more thorough understanding of
the relation between closed and open structures to the proposal of a
new measure that captures a topological configuration at the interface
between the two types of structure. The idea was to identify brokerage
positions in which a node can intermediate between otherwise
disconnected cohesive groups of contacts \cite{vedres_2010}. In such
cases, the node's local network can be seen as both open and closed:
open in that it is rich in structural holes separating distinct groups
of contacts; and closed in that it is at the same time rich in
third-party relationships within each of the groups with which the
node is affiliated.

In qualitative agreement with the organisational literature on
Simmelian ties
\cite{dekker_2006,krackhardt_1998,krackhardt_1999,krackhardt_2002,tortoriello_2010},
we proposed to call Simmelian brokerage the new measure for detecting
such structural positions. Simmelian brokerage helps differentiate
between brokerage positions of nodes with the same degree and the same
local clustering coefficient, but with a different configuration of
links in their local neighbourhoods. In those cases, effective size
would also remain unchanged as, all else being equal, it is not
sensitive to variations in the positions of links. However, brokerage
opportunities are likely to differ when, simply by reshuffling the
same number of links across a node's local neighbourhood, there is a
variation in the number of socially cohesive groups with which the
node is affiliated. Simmelian brokerage, unlike effective size, is
sensitive precisely to these variations in group affiliation that
result from a change in the position of links.

Our findings can nourish the theoretical debate over the relative
salience of closed and open structures for social capital, and will
inform further research on the generative mechanisms of social
capital. On the one hand, empirical tests of the relative benefits of
closed and open structures will now find in our proposed relation
between clustering and effective size a sound argument safeguarding
against problems of multicollinearity, typically arising as a result
of the simultaneous inclusion of both measures as explanatory
variables in multivariate regression models. On the other, future
research on the relative benefits of cohesive and brokered networks
will benefit from the application of Simmelian brokerage to a number
of empirical domains. In this sense, our study will help reconcile the
apparently opposing results that various strands of literature have
uncovered on the structural foundations of social capital
\cite{aral_2011,ahuja_2000,fleming_2007,gargiulo_2000,ingram_2000,lingo_2010,reagans_2003}.

Simmelian brokerage, as a new topological measure of network
structure, can also spur a wealth of research broadly concerned with
the topology and dynamics of complex networks. For the sake of
simplicity, in this paper we have restricted our focus primarily to
the case of unweighted networks. However, as was sketched out in
Section~\ref{sec:6}, the relation between clustering coefficient and
effective size can easily be generalised to the case of weighted
graphs, which will in turn enable Simmelian brokerage to be also
extended to weighted graphs.

More generally, the main implication of our study for research on
complex networked systems lies in the change of perspective entailed
by our emphasis on structural cleavages, as opposed to ties, that we
borrowed from the burgeoning network literature in the social
sciences. In this sense, our study may suggest a number of possible
and previously neglected ways in which, simply by deflecting attention
from the presence to the absence of a tie, new insights can be gained
on the organisation, functioning and dynamics of a variety of systems.


\begin{thebibliography}{99}

%
%


\bibitem{ahuja_2000}
Ahuja, G.: Collaboration networks, structural holes, and innovation: A longitudinal study. Administrative Science Quarterly. 45, 425-455 (2000).

\bibitem{aral_2011} 
Aral, S., Van Alstyne, M.: The diversity-bandwidth trade-off. American Journal of Sociology. 117(1), 90-171 (2011).

\bibitem{Barrat2004} 
Barrat, A., Barthélemy, M., Pastor-Satorras, R. and Vespignani, A.: The architecture of complex weighted networks. Proceedings of the National Academy of Sciences of the United States of America. 101, 3747-3752 (2004).

\bibitem{baum_2012}
Baum, J.A.C., McEvily, B., Rowley, T.J.: Better with age? Tie longevity and the performance Implications of bridging and closure. Organization Science. 23, 529-546 (2012).

\bibitem{borgatti97} 
Borgatti, S.P.: Structural holes: Unpacking Burt's redundancy measures. Connections. 20, 35 (1997).

\bibitem{brass_1995}
Brass, D.J.: It's all in your social network. In: Ford, C.M., Gioia, D.A. (eds.), Creative Action in Organizations, pp. 94-98. Sage, Thousand Oaks, CA (1995).

\bibitem{burt_1992}
Burt, R.S.: Structural Holes. The Social Structure of Competition. Harvard University Press, Cambridge MA (1992).

\bibitem{burt_2004}
Burt, R.S.: Structural holes and good ideas. American Journal of Sociology. 110, 349-399 (2004).

\bibitem{burt_2005}
Burt, R.S.: Brokerage and Closure. Oxford University Press, Oxford (2005).

\bibitem{burt_2010}
Burt, R.S.: Neighbor Networks. Oxford University Press, Oxford (2010).

\bibitem{burt_1998}
Burt, R. S.: The gender of social capital. Rationality and Society. 10, 5-46 (1998).

\bibitem{burt_1995}
Burt, R. S., and Knez, M.: Kinds of third-party effects on trust. Rationality and Society. 7, 255-292 (1995).

\bibitem{centola_2007}
Centola, D., Macy, M.W.: Complex contagion and the weakness of long ties. American Journal of Sociology. 113, 702-734 (2007).

\bibitem{coleman_1988}
Coleman, J. S.: Social capital in the creation of human capital. American Journal of Sociology. 94, S95-S120 (1988). 

\bibitem{coleman_1990}
Coleman, J.S.: Foundations of Social Theory. Harvard University Press, Cambridge, MA (1990).

\bibitem{cover} 
Cover T.M., Thomas J. A.: Elements of Information Theory. Wiley, New York (1991).

\bibitem{davis_1970}
Davis, J.A.: Clustering and hierarchy in interpersonal relations: Testing two graph theoretical models on 742 sociomatrices. American Sociological Review. 35(5), 843-851 (1970).

\bibitem{davis_1971}
Davis, J.A., Holland, P.W., Leinhardt, S.: Comments on Professor Mazur's hypothesis about interpersonal sentiments. American Sociological Review. 36, 309-311 (1971).

\bibitem{dekker_2006}
Dekker, D.: Measures of Simmelian tie strength, Simmelian brokerage, and Simmelianly brokered. Journal of Social Structure. 7(1), 1-22 (2006).

\bibitem{fleming_2007}
Fleming, L., Mingo, S., Chen, D.: Collaborative brokerage, generative creativity, and creative success. Administrative Science Quarterly. 52, 443-475 (2007).

\bibitem{fortunato_2010}
Fortunato, S.: Community detection in graphs. Physics Reports. 486, 75-174 (2010).

\bibitem{friedkin_2004}
Friedkin, N.E.: Social Cohesion. Annual RevIew of Sociology. 30, 409-25 (2004).

\bibitem{gargiulo_2000}
Gargiulo, M., Benassi, M.: Trapped in your own net? Network cohesion, structural holes, and the adaptation of social capital. Organization Science. 11: 183-196 (2000).

\bibitem{gargiulo_2009}
Gargiulo, M., Ertug, G., Galunic, C.: The two faces of control: Network closure and individual performance among knowledge workers. Administrative Science Quarterly. 54, 299-333 (2009).

\bibitem{gould_1991}
Gould, R.V.: Multiple networks and mobilization in the Paris Commune, 1871. American Sociological Review. 56, 716-729 (1991).

\bibitem{granovetter_1973} 
Granovetter, M.: The strength of weak ties. American Journal of Sociology. 78, 1360-1380 (1973).

\bibitem{granovetter_2005}
Granovetter, M.: The impact of social structure on economic outcomes. Journal of Economic Perspectives. 19(1), 33-50 (2005).

\bibitem{hansen_1999} 
Hansen, M.T.: The search-transfer problem: The role of weak ties in sharing knowledge across organization subunits. Administrative Science Quarterly. 44, 232-248 (1999).

\bibitem{holland_1970}
Holland, P.W., Leinhardt, S.: A method for detecting structure in sociometric data. American Journal of Sociology. 76, 492-513 (1970).

\bibitem{holland_1971}
Holland, P.W., Leinhardt, S.: Transitivity in structural models of small groups. Comparative Group Studies. 2, 107-124 (1971).

\bibitem{Holme2002}
Holme, P. Kim, B. J.: Growing scale-free networks with tunable clustering. Phys. Rev. E. 65, 026107 (2002).

\bibitem{Holme2007} 
Holme, P., Park, S. M., Kim, B. J. and Edling, C. R.: Korean university life in a network perspective: Dynamics of a large affiliation network. Physica A: Statistical Mechanics and its Applications. 373, 821-830 (2007).

\bibitem{ingram_2000}
Ingram, P., Roberts, P.W.: Friendships among competitors in the Sydney hotel industry. American Journal of Sociology. 106(2), 387-423 (2000).

\bibitem{krackhardt_1998}
Krackhardt, D.: Simmelian tie: Super strong and sticky. In: R. M. Kramer, R.M., Neale, M.A. (eds.), Power and influence in organizations, p. 21-38. Sage, Thousand Oaks, CA (1998).

\bibitem{krackhardt_1999}
Krackhardt, D.: The ties that torture: Simmelian tie analysis in organizations. Research in the Sociology of Organizations. 16, 183-210 (1999).

\bibitem{krackhardt_2002}
Krackhardt, D., Kilduff, M.: Structure, culture and Simmelian ties in entrepreneurial firms. Social Networks. 24, 279-290 (2002).

\bibitem{lambiotte_2009}
Lambiotte, R., Panzarasa, P.: Communities, knowledge creation, and information diffusion. Journal of Informetrics. 3(3), 180-190 (2009). 

\bibitem{lm01}  
Latora, V., Marchiori, M.: Efficient behavior of small-world networks. Phys. Rev. Lett. 87, 198701 (2001). 

\bibitem{lm03} 
Latora, V., Marchiori, M.: Economic behavior of small-world networks. Eur. Phys. J. B. 32, 249-263 (2003).

\bibitem{lin_2001}
Lin, N.: Social Capital. A Theory of Social Structure and Action. Cambridge University Press, New York (2001).

\bibitem{lin_2001a}
Lin, N., Cook, K., Burt, R.S. (eds.): Social Capital. Theory and Research. Aldine Transaction, New Brunswick and London (2001).

\bibitem{lingo_2010}
Long Lingo, E., O'Mahony, S.: Nexus work: Brokerage on creative projects. Administrative Science Quarterly. 55, 47-81 (2010).

\bibitem{luce_1949}
Luce, R.D., Perry, A.D.: A method of matrix analysis of group structure. Psychometrika. 14(1), 95-116 (1949).

\bibitem{mizruchi_2001}
Mizruchi, M., Stearns, L.B.: Getting deals done: The use of social networks in bank decision-making. American Sociological Review. 66, 647- 471 (2001).

\bibitem{nahapiet_1998}
Nahapiet, J., Ghoshal, S.: Social capital, intellectual capital, and the organizational advantage. Academy of Management Review. 23, 242-266 (1998).

\bibitem{obstfeld_2005}
Obstfeld, D.: Social networks, the \textit{tertius iungens} orientation, and involvement in innovation. Administrative Science Quarterly. 50, 100-130 (2005).

\bibitem{Onnela2005} 
Onnela, J.-P., Saram\"aki, J., Kert\'esz, J. and Kaski, K.: Intensity and coherence of motifs in weighted complex networks. Phys. Rev. E. 71, 065103 (2005).

\bibitem{opsahl_2009}
Opsahl, T., Panzarasa, P.: Clustering in weighted networks. Social Networks. 31, 155-163 (2009).

\bibitem{palla_2005}
Palla, G., Der\'enyi, I., Farkas, I., Vicsek. T.: Uncovering the overlapping community structure of complex networks in nature and society.
Nature. 435 (7043), 814-818 (2005).

\bibitem{perry-smith_2006} 
Perry-Smith, J.E.: Social yet creative: The role of social relationships in facilitating individual creativity. Academy of Management Journal. 49, 85-101 (2006).

\bibitem{podolny_1997}
Podolny, J. M., Baron, J. N.: Resources and relationships: Social networks and mobility in the workplace. American Sociological Review. 62, 673-693 (1997).

\bibitem{Ravasz2003}
Ravasz, E., Barab\'asi, A.-L.: Hierarchical organization in complex
networks. Phys. Rev. E 67, 026112 (2003).

\bibitem{reagans_2003}
Reagans, R., McEvily, B.: Network structure and knowledge transfer: The effects of cohesion and range. Administrative Science Quarterly. 48, 240-267 (2003).

\bibitem{reagans_2001}
Reagans, R., Zuckerman, E.: Networks, diversity and performance: The social capital of R\&D units. Organization Science. 12, 502-517 (2001).

\bibitem{rodan_2004}
Rodan, S., Galunic, C.: More than network structure: How knowledge heterogeneity influences managerial performance and innovativeness. Strategic Management Journal, 25, 541-562 (2004).

\bibitem{Saramaki2007} 
Saram\"aki, J., Kivel\"a, M., Onnela, J.-P., Kaski, K. and Kert\'esz, J.: Generalizations of the clustering coefficient to weighted complex networks. Phys. Rev. E. 75, 027105 (2007).

\bibitem{Serrano2005} Serrano, M.A., Bogu\~n\'a, M.: Tuning clustering in random networks with arbitrary degree distributions. Phys. Rev. E. 72, 036133 (2005).

\bibitem{simmel_2011}
Simmel, G.: The Sociology of Georg Simmel. Trans. by K.H. Wolff.  Lightning Source, Milton Keynes UK [1923] (2011).

\bibitem{sosa_2011}
Sosa, M.E.: Where do creative interactions come from? The role of tie content and social networks. Organization Science. 22, 1-21 (2011).

\bibitem{stovel_2012}
Stovel, K., Shaw, L.:  Brokerage. Annual Review of Sociology. 38, 139-158 (2012).

\bibitem{tortoriello_2010}
Tortoriello, M., Krackhardt, D.: Activating cross-boundary knowledge: The role of Simmelian ties in the generation of innovations. Academy of Management Journal. 53(1), 167-181 (2010).

\bibitem{tortoriello_2012}
Tortoriello, M., Reagans, R., McEvily, B.: Bridging the knowledge gap: The influence of strong ties, network cohesion, and network range on the transfer of knowledge between organizational units. Organization Science. 23, 1024-1039 (2012).

\bibitem{uzzi_1997} 
Uzzi, B.: Social structure and competition in interfirm networks: The paradox of embeddedness. Administrative Science Quarterly. 43, 35-67 (1997).

\bibitem{uzzi_2005} 
Uzzi, B., Spiro, J.: Collaboration and creativity: The small world problem. American Journal of Sociology. 111, 447-504 (2005).

\bibitem{Vazquez2002} V\'azquez, A., Pastor-Satorras, R., Vespignani, A.: Large-scale topological and dynamical properties of the Internet. Phys. Rev. E. 65, 066130 (2002).

\bibitem{Vazquez2003} V\'azquez, A.: Growing network with local rules: Preferential attachment, clustering hierarchy, and degree correlations. Phys. Rev. E. 67, 056104 (2003).

\bibitem{vedres_2010}
Vedres, B., Stark, D.: Structural folds: Generative disruption in overlapping groups. American Journal of Sociology. 115(4), 1150-1190 (2010).

\bibitem{wattsbook} 
Watts, D.J.: Small Worlds: The Dynamics of Networks between Order and Randomness. Princeton University Press, Princeton, New Jersey (1999).

\bibitem{ws98}  
Watts, D.J., Strogatz, S.H.: Collective dynamics of small-world networks. Nature. 393, 440 (1998).

\bibitem{Zhang2005}
Zhang, B., Horvath, S.: A general framework for weighted gene co-expression network analysis. Stat. App. Genet. Mol. Biol. 4, 17 (2005).


\end{thebibliography}
\end{document}